\documentstyle[aps,prb,twocolumn,epsf]{revtex}
\begin{document}
\draft

\preprint{IUCM96-022} 
\title{Plasmon Modes and Correlation Functions in Quantum Wires and
Hall Bars}
 
\author{U.\ Z\"ulicke and A.\ H.\ MacDonald}

\address{
Department of Physics, Indiana University, Bloomington, Indiana 47405,
U.S.A.}

\date{\today}
\maketitle

{\tightenlines
\begin{abstract}

We present microscopic derivations of the one-dimensional low-energy
boson effective Hamiltonians of quantum wire and quantum Hall bar
systems.  The quantum Hall system is distinguished by its spatial
separation of oppositely directed electrons.  We discuss qualitative
differences in the plasmon collective mode dispersions and the ground
state correlation functions of the two systems which are consequences
of this difference.  The slowly-decaying quasi-solid correlations
expected in a quantum wire are strongly suppressed in quantum Hall
bar systems. 

\end{abstract}
}
 

\narrowtext

\section{Introduction}
The quantum Hall effect occurs in two-dimensional (2D) electron
systems (ES) when the chemical potential lies in a charge gap which
occurs at a density ($n^{*}$) which is dependent on magnetic field
($B$). The $B$ dependence of $n^{*}$ requires\cite{ahm:braz:96}
gapless excitations localized at the edge of the 2D ES. The low-energy
effective Hamiltonian which describes this edge system is simplest
when the edge is sharp\cite{smoothedge} on a microscopic length scale
and the bulk Landau level filling factor $\nu^{*} = 2 \pi \ell^2
n^{*} =1/m$ with $m$ being an odd integer. In this case, the
low-energy edge excitations can be mapped to those of a
one-dimensional (1D) fermion system\cite{bih:prb:82,ahm:prl:90} and
described\cite{wen:prb:90,wen:int:92} by a version of the
Tomonaga-Luttinger (TL) model for 1D fermion
systems\cite{emery,sol:adv:79,fdmh:jpc:81} modified to account for
the magnetic field and the long-range of the electron-electron
interaction. Edge excitations in quantum Hall (QH) bars are analogous
to the excitations of electron systems in quantum
wires\cite{somerecentrefs} which are also described at low energies
by a TL model modified to account for long-range interactions. Close
relationships exist between studies of the effect of Coulomb
interactions on the transport  properties of QH
systems\cite{oreg:prl:95,newmoon} and analogous studies of quantum
wires.\cite{fab:prl:94,gia:prb:95} There are, however, important
distinctions between these two systems which result from the spatial
separation of oppositely directed electrons in the QH case and are
the subject of this paper. As we explain below, the energy-wavevector
relationship for the plasmon boson states of quantum wires and QH
bars have quite different microscopic underpinnings. In addition, the
strong quasi-solid correlations expected\cite{hjs:prl:93} in a
quantum wire are suppressed in typical Hall bar
systems.\cite{brey:95}

\section{Non-Interacting Electrons}
The analogy between quantum wires and quantum Hall bars is most
direct for $\nu=1$ and we begin by discussing this case, first for
non-interacting electrons. To form a quantum wire, electrons in a
2D ES are confined in an additional direction, say the
$\hat y$-direction, while the motion in the remaining
($\hat x$-)direction stays free. With periodic boundary conditions
applied over a length $L$ in the $\hat x$-direction, the electronic
single-particle wave functions then have the form 
\begin{equation}\label{wavefunct}
\psi^{\mbox{\tiny QW}}_{n, k} (x, y) = \frac{1}{\sqrt{L}} \,
e^{i k x} \,\, \chi^{\mbox{\tiny QW}}_{n}(y)
\end{equation}
where $\chi^{\mbox{\tiny QW}}_{n}(y)$ is the $n$-th discrete subband
state for the quantum wire. The single-particle eigenenergy
$\varepsilon_{n,k} = E_n +  \hbar^2 k^2 / 2 m^{*}$. For wires widths
comparable to or smaller than the typical distance between electrons,
the energy spacing between different subbands will be larger than the
Fermi energy and only the lowest ($n = 0$) subband will be occupied
in the ground and low-energy excited states. The ground state of the
non-interacting many-electron system is a 1D Fermi gas state in which
all single-particle states with $n = 0$ and $|k| \le k_F$ are
occupied. The subband wavefunction leads only to form factors which
will modify the electron-electron interaction in the system at short
distances. The situation for a 2D ES in a strong perpendicular
magnetic field is similar. For a Hall bar geometry where the system
is confined in the $\hat y$-direction and periodic boundary
conditions are applied in the $\hat x$-direction, the Landau gauge
[$\vec A = (-By, 0, 0)$] wavefunctions have the form\cite{caveatsf} 
\begin{equation}
\psi^{\mbox{\tiny HB}}_{n, k} (x, y) = \frac{1}{\sqrt{L}} \, e^{i k x}
\,\, \chi^{\mbox{\tiny HB}}_{n}(y - \ell^2 k)\mbox{ .}
\end{equation}
Here $\ell:=[\hbar c/(e B)]^{1/2}$ is the magnetic length, $n$ is the
Landau level index, and $\chi^{\mbox{\tiny HB}}_{n}(y)$ is the wave
function of a harmonic oscillator with frequency $\omega_c = e B/(
m^* c)$ which is localized to a length $\sim \ell$ around $y=0$. The
oscillator wavefunctions play the role of the subband wavefunction in
a quantum wire, but in the Hall bar case they are displaced from the
origin by a distance proportional to wave vector $k$. In addition,
the dependence of the single-particle eigenenergy, $\varepsilon_{n,k}
= \hbar \omega_c (n+1/2) + V^{\mbox{\tiny ext}}(\ell^2 k)$, on $k$ is
due to the confinement potential $V^{\mbox{\tiny ext}}(y)$ rather
than to the kinetic energy. In the strong magnetic field limit, only
$n=0$ states will be occupied at $\nu =1$, even when the width of the
Hall bar is macroscopic, and states at the Fermi energy with $k = \pm
k_F$ will be localized at opposite edges of the sample, as
illustrated in Fig.~\ref{wirebar}.  This property of Hall bar systems
plays the central role role in the edge state picture of the integer
quantum Hall effect for non-interacting
electrons.\cite{bih:prb:82,butt:prb:88} The sample width $W = 2 k_F
\ell^2$ is assumed to be much larger than the magnetic length $\ell$
throughout this paper. When this condition is not satisfied the
distinction between a quantum wire and a quantum Hall bar blurs. 

\section{Coulomb Matrix Elements and Low-Energy Hamiltonian}
\begin{figure}
\epsfxsize3.7in
\centerline{\epsffile{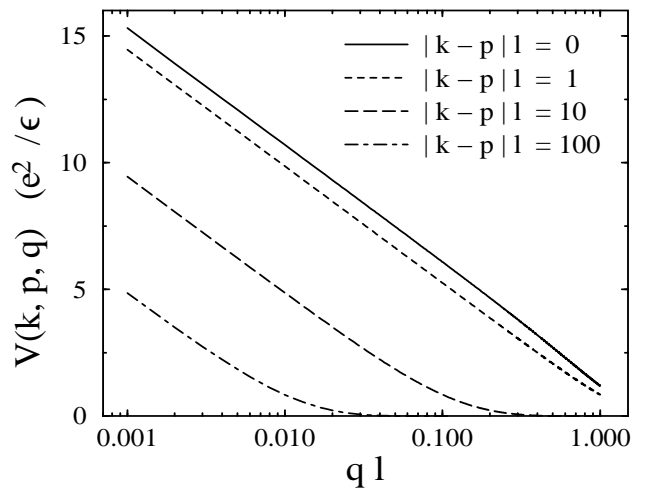}}
\caption{Matrix element Eq.~(\ref{intmatel}) evaluated for Coulomb
interaction. Three regimes can be distinguished if $q \ell < 1$.
For the case $|k - p| \ell < 1$ we find that $V(k, p, q) \sim -\ln
(q \ell)$ and is essentially independent of $k, p$. For $|k - p| \ell
> 1$ and $q |k-p| \ell^2 < 1$, it is found that $V(k, p, q) \sim
-\ln(q |k-p| \ell^2)$. Finally, the matrix element is negligibly
small for $q |k-p| \ell^2 > 1$.}
\label{coulmat}
\end{figure}

Quantum wire and QH bar systems are described by microscopic
Hamiltonians of the same form
\begin{mathletters}\label{microscop}
\begin{eqnarray}
H  &=& H^{0} + H^{\mbox{\tiny int}} \\
H^{0} &=& \sum_{k} \, \hbar (|k|-k_F)\, v_{\mbox{\tiny F}} \,
c_{k}^\dagger c_k \\ \label{interact}
H^{\mbox{\tiny int}} &=& \frac{1}{2L} \sum_{k, p, q} V(k, p, q) \,\,
c_{k+q}^\dagger c_{p}^\dagger c_{p+q} c_k
\end{eqnarray}
\end{mathletters}
where $H^{0}$ is the one-body term in the Hamiltonian and the 
single-particle energy has been linearized around $k = \pm k_F$ so
that $v_{\mbox{\tiny F}} = \hbar k_F / m^{*}$ in the quantum wire
case while 
\begin{figure}[b]
\epsfxsize3.5in
\centerline{\epsffile{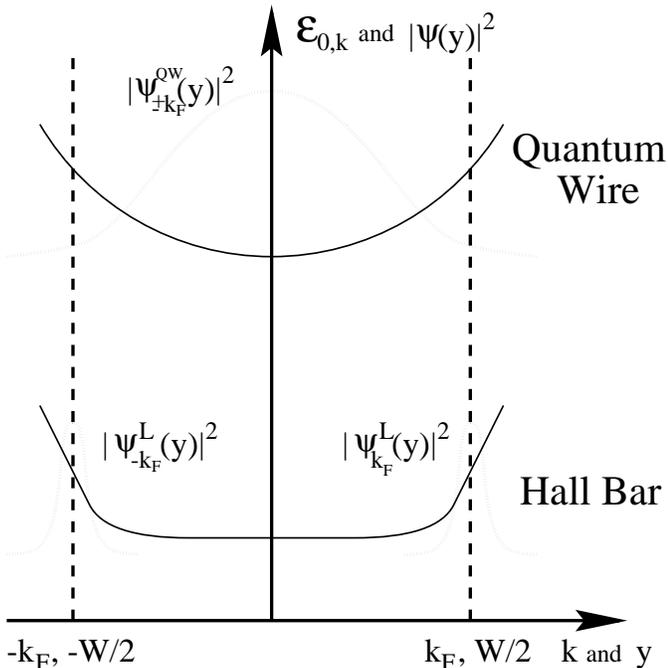}}
\caption{Comparison of quantum wire and quantum Hall bar systems.
Both the dependence of the single-particle energy $\varepsilon_{0,
k}$ on wave vector $k$ (solid lines) and the spatial extent of the
lateral part of the wave function [$\equiv | \psi_{0,k}(y) |^{2}$
depicted as dotted lines] are shown schematically. The principal
difference is the strong localization of $| \psi^{HB}_{0,\pm k_F}
(y)|^2$ at opposite edges of the sample in the Hall bar case,
compared to the identity of $|\psi^{QW}_{0,\pm k_F} (y)|^2$ in the
quantum wire case.}
\label{wirebar}
\end{figure}
\begin{equation}\label{fermvel}
v_{\mbox{\tiny F}} = \frac{\ell^2}{\hbar} \left. \frac{d
V^{\mbox{\tiny ext}}}{d y} \right|_{W/2}
\end{equation}
in the Hall bar case. The $y$-dependent part of the wavefunctions
enters crucially into the form of the matrix element $V(k, p, q)$.
For quantum wires,\cite{dassarma,gold:prb:90} the function
$\chi^{\mbox{\tiny QW}}_{n}(y)$ does not depend on $k$, and the
interaction matrix element is a function of momentum transfer $q$
only. Up to an irrelevant constant, it is then possible to rewrite
Eqs.~(\ref{microscop}) as
\begin{equation}\label{1Dhamilt}
H_{\mbox{\tiny eff}}^{\mbox{\tiny QW}} = H^{0} + \frac{1}{2 L}
\sum_{q \ne 0} V^{\mbox{\tiny QW}}_q \varrho_q \varrho_{-q} \mbox{ .}
\end{equation}
Here $\varrho_q = \sum_k c_{k+q}^{\dagger} c_k$ is the 1D Fourier
transform of the density operator. For Coulomb interaction, the
effective 1D potential at small $q$ is\cite{dassarma,gold:prb:90} 
\begin{equation}
V^{\mbox{\tiny QW}}_q = \frac{e^2}{\epsilon } (-2) \ln{\left[\alpha
\, q\, W \right]}
\end{equation}
where $W$ is the width of the quantum wire and $\alpha$ is a constant 
of order unity which depends on the details of the confining
potential. In the QH bar, however, single-particle states with
different relative momentum are separated in the $\hat y$-direction.
(See Fig.~\ref{wirebar}.) Consequently, the matrix element
$V(k, p, q)$ depends on both $q$ and $k-p$. If the originally 2D
interaction is $U(\vec r)$ and has the 2D Fourier transform
$U(\vec q) = U(q_x, q_y)$, we find that 
\begin{equation}\label{intmatel}
V(k, p, q) = \frac{e^{-\frac{1}{2}(q\ell)^2}}{\ell }\,\,
\int_{-\infty}^{\infty} d\kappa \,\, e^{-\frac{1}{2} \kappa^2}\,
U(q, \kappa \ell^{-1})\, e^{i\kappa (k-p) \ell}.
\end{equation}
For the physically relevant Coulomb interaction in the limit of small
momentum transfer $q\le\ell^{-1}$ we obtain (see also
Fig.~\ref{coulmat})
\begin{equation}\label{coulomb}
V(k, p, q) = \frac{e^2}{\epsilon } \left\{
\begin{tabular}{ll}
$2\,\mbox{K}_0(|q (k-p) \ell^2|)$ & $\mbox{for } |k-p| > \ell^{-1}$\\
$-2\ln{\left[\sqrt{\frac{\gamma}{8}} \, q\ell\right]}$ & $\mbox{for }
|k-p| \le \ell^{-1}$
\end{tabular}
\right.
\end{equation}
where $\epsilon$ is the host semiconductor dielectric constant,
and $\gamma \sim 1.78$ is the exponential of Euler's constant. 
Corrections to Eq.~(\ref{coulomb}) are analytic in $q$, in
$|k-p|^{-1}$ for large $|k-p|$, and in $|k-p|$ for small $|k-p|$ and
are negligible for our purposes. Similar expressions for two-particle
Coulomb matrix elements in a Hall bar have been reported
previously.\cite{lee:prb:90,wen:prb:91a,bih:prl:93,wen:prb:94}
Here we want to use this expression to derive a 1D effective
Hamiltonian similar to Eq.~(\ref{1Dhamilt}), to describe the
low-energy excitations of a Hall bar. It is useful to separate the
$q=0$ term in the Hamiltonian which, in contrast to the quantum wire
case, is not a constant. Defining $\tilde{V}(k-p) := V(k, p, 0)$ we
find that 
\begin{equation}
H^{\mbox{\tiny HB}} = H^{0} + H^{q=0} + H^{\mbox{\tiny TL}}
\end{equation}
where 
\begin{mathletters}
\begin{eqnarray}
H^{q=0} &=& \frac{1}{2 L}\sum_{k, p} \tilde{V}(k-p)c_{k}^\dagger
c_{p}^\dagger c_p c_k \\ &\simeq& \frac{1}{L} \sum_{k}\left[
\sum_{p=-k_F}^{k_F} \tilde{V}(k - p) \right]\, c_{k}^\dagger c_k
\end{eqnarray}
\end{mathletters}
The last line holds in the low-energy sector of the Hamiltonian where
number operator fluctuations are negligible except for
single-particle states close to the edge of the system which do not
contribute importantly to the sum. $H^{q=0}$ simply adds the
electrostatic (Hartree) contribution to the single-particle energies
which would be present in a Hartree self-consistent field theory.
This contribution is an irrelevant constant for a quantum wire but is
state dependent in a Hall bar. The Hartree energy is positive and
smaller in magnitude for states with $|k| > k_F$ since they are
localized farther from the other electrons. Linearizing the Hartree
single-particle energy we find that for low energies and $W \gg \ell$ 
\begin{equation}\label{elstatic}
H^{q=0}_{\mbox{\tiny eff}} = \sum_{k} \left(-\frac{e^2}{\epsilon \pi}
\right) \, (|k| - k_F) \, \ln{\left[\sqrt{2\gamma} \frac{W}{\ell}
\right]} c_{k}^{\dagger} c_k \mbox{ .}
\end{equation}

The low-energy physics of a 1D ES can be captured in an approximation
where the Hamiltonian is projected onto sectors where the number of
right going fermions $\big( N_R := \sum_{k>0} \big< c_k^\dagger c_k
\big> \big)$ and the number of left going fermions $\big( N_L :=
\sum_{k<0} \big< c^\dagger_kc_k \big> \big)$ are fixed. As is well
known from studies of TL models\cite{emery,sol:adv:79,fdmh:jpc:81}
for 1D ES, this projection permits one-body terms linearized around
the Fermi points to be expressed in terms of density operators
$\varrho^{\mbox{\tiny R,L}}_q := \sum_{k > 0 \atop k < 0} 
c^\dagger_{k + q} c_k$. $\big(\varrho_q = \varrho^{\mbox{\tiny L}}_q
+ \varrho^{\mbox{\tiny R}}_q \big)$ Using this procedure,
$H^{q=0}_{\mbox{\tiny eff}}$ can be lumped with the $q \ne 0$
interaction terms in the Hamiltonian to obtain   
\begin{eqnarray}\label{1Deffhamilt}
H^{\mbox{\tiny HB}}_{\mbox{\tiny eff}} &=& H^{0} + \frac{1}{L}
\sum_{q>0} \left\{ V^{\mbox{\tiny intra}}_q \left[
\varrho^{\mbox{\tiny L}}_q \varrho^{\mbox{\tiny L}}_{-q} +
\varrho^{\mbox{\tiny R}}_q \varrho^{\mbox{\tiny R}}_{-q} \right]
\right. \nonumber \\ && \hspace{2cm} + \left.
 V^{\mbox{\tiny inter}}_q \left[ \varrho^{\mbox{\tiny L}}_q
\varrho^{\mbox{\tiny R}}_{-q} + \varrho^{\mbox{\tiny R}}_q
\varrho^{\mbox{\tiny L}}_{-q} \right] \right\}
\end{eqnarray}
where the effective 1D intra-edge and inter-edge interactions are 
$$V^{\mbox{\tiny intra}}_q := \frac{e^2}{\epsilon } (-2) \ln{\left[
\frac{\gamma}{2} \, q\, W \right]} \mbox{ and }
V^{\mbox{\tiny inter}}_q := \frac{e^2}{\epsilon } 2 \mbox{K}_0(q W)
\mbox{ .}$$
Microscopically, the $q \ne 0$ terms in the interaction Hamiltonian
represent the loss of exchange energy when a density wave is created
in the system.  To obtain this result we have used the fact that the
dependence of $V(k,p,q)$ on $k$ and $p$ is negligible at small $q$
when $k$ and $p$ are near the same Fermi point and appealed to
linearization in setting $|k-p| = W/\ell^2$ when $k$ and $p$ are near
opposite Fermi points.

Previous studies addressing the effect of Coulomb interaction in QH
systems have used a Hamiltonian of the form shown in
Eq.~(\ref{1Deffhamilt}) as starting point. In this work the
Hamiltonian implicitly or explicitly contains two undetermined
parameters: the bare Fermi velocity $v_{\mbox{\tiny F}}$ appearing in
$H_0$ and a cut-off length, generally assumed to be microscopic,
appearing in the expression for $V^{\mbox{\tiny intra}}_q$. In our
microscopic analysis, the length appearing in
$V^{\mbox{\tiny intra}}_q$ is the macroscopic sample width $W$ and
$v_{\mbox{\tiny F}}$, defined by Eq.~(\ref{fermvel}), is dependent on
the external potential. However, it is important to realize that the
projection onto fixed $N_R$ and $N_L$ sectors obviates the
distinction we have made between one-body and two-body terms in the
underlying microscopic Hamiltonian. The identification of a bare
Fermi velocity associated with the one-body term plays {\em no role}
in the physics. In our analysis, the intra-edge interaction
represents the sum of terms originating from the one-body Hartree
energy which leads to an attractive effective interaction and the
$q \ne 0 $ terms in the intra-edge interaction which are repulsive.
We could as well have grouped the Hartree term with the one-body term
in the Hamiltonian. With this choice, the one-body term would vanish
if the external potential originated from a positively-charged
background which precisely cancelled the ground state electron charge
density. In typical experimental situations the external potential
which attracts electrons to the Hall bar is weaker near the edge than
in the neutralizing background model frequently used in theoretical
model calculations so that the Fermi velocity is negative when the
Hartree term is grouped with the one-body terms. The negative Fermi
velocity needn't have any physical consequences, however, since in
the case of interest the $q \ne 0$ terms in the Hamiltonian stabilize
all excitations.
For smoother edges this Fermi velocity becomes more
and more negative and eventually the system will become unstable to
edge reconstructions,\cite{wen:prb:94,ahm:aust:93,edgerecon}
signaling a phase transition to a state with a different and more
complicated low-energy effective Hamiltonian. However, this
instability involves `ultraviolet' physics which is beyond the scope
of the present study.

We see from the expression for $V^{\mbox{\tiny inter}}_q$ that the
inter-edge interaction is important only if $q W < 1$, because the
modified Bessel function K$_0(x)$ decays rapidly for $x > 1$.
Physically, the Coulomb potential due to a 1D density wave is weak
when viewed from a point removed from the 1D system by a distance
longer than the period of the wave. For small $x$, K$_0(x) \to -
\ln{\left[\frac{\gamma}{2} x \right]}$ so that in the limit $q \ll
W^{-1}$, we have $V^{\mbox{\tiny intra}}_q = V^{\mbox{\tiny inter}}_q
= V^{\mbox{\tiny QW}}_q$. The contribution to the Hamiltonian of a
quantum Hall bar from modes with a wavelength exceeding $W$ is
identical to the corresponding contribution to the effective 1D
Hamiltonian for the quantum wire.

\section{Bosonization}
TL models described by a Hamiltonian displayed in
Eq.~(\ref{1Deffhamilt}) can be solved by means of
bosonization.\cite{emery,sol:adv:79,fdmh:jpc:81} This is possible
because within the restricted Hilbert space of low-energy, small-$q$
excitations around a uniform ground state, the density operators
$\varrho^{\mbox{\tiny L,R}}_q$ obey simple bosonic commutation
relations. If $\nu$ denotes the occupation number of single-particle
states in the uniform many-particle ground state around which the
excitations occur, the commutation relations are
\begin{mathletters}\label{anomaly}
\begin{eqnarray}
\left[ \varrho^{\mbox{\tiny L}}_q, \varrho^{\mbox{\tiny L}}_{q'}
\right] &=& \nu \, \frac{q L}{2 \pi} \, \delta_{q + q' , 0} \\
\left[ \varrho^{\mbox{\tiny R}}_q, \varrho^{\mbox{\tiny R}}_{q'}
\right] &=& - \nu \, \frac{q L}{2 \pi} \, \delta_{q + q' , 0} \\
\left[ \varrho^{\mbox{\tiny L}}_q, \varrho^{\mbox{\tiny R}}_{q'}
\right] &=& 0
\end{eqnarray}
\end{mathletters}
So far, we have only considered the case of $\nu = 1$, which is the
generic case for a quantum wire. Below we identify $\nu$ as the
Landau level filling factor of the QH bar and comment on the
validity of the Luttinger liquid description of QH
edges\cite{wen:prb:90} at $\nu < 1$.

In order to understand the intriguing features which are special to
the physics of a 1D ES, it has proven to be useful\cite{fdmh:jpc:81}
to introduce two new bosonic fields $\theta(x)$ and $\phi(x)$, called
{\em phase fields}. In what follows, we assume that $N_R$ and $N_L$
are fixed. It turns out that the Hamiltonian (\ref{1Deffhamilt}) can
then be rewritten (uniquely up to a constant) as a quadratic
Hamiltonian in these phase fields. Here, we have an additional
motivation for following this procedure; the generalization of the TL
picture to fractional filling factors $\nu$ is straightforward once
the phase fields are introduced. The theory can be formulated in the
phase field formalism for an arbitrary filling factor, and the value
of $\nu$ enters the theory only when calculating physical observables
like densities and currents. Before going into algebraic details, we
comment on the justification of the TL model for fractional QH edges.
(For a related discussion see Ref.~\onlinecite{mpaf:int:94}.) The
explicit derivation for a quantum Hall bar outlined in the previous
section does not generalize to the $\nu < 1$ case. However, using
arguments based on the analytical structure of many-body wave 
functions that describe the 2D ES in the fractional QH regime, it can
be shown that for abrupt edges the one-to-one correspondence between
the low-lying excitations in this system and the excitations of a 1D
boson system holds\cite{ahm:prl:90,ahm:braz:96} for $\nu =1/m$ where
$m$ is an odd integer. The form of the theory is essentially
fixed\cite{ahm:braz:96} by this bosonization property and the
requirement that the theory recover the fractional quantum Hall
effect under appropriate circumstances. Therefore we believe that,
for abrupt edges, the low-energy Hamiltonian for a QH bar can be
written in the form of Eq.~(\ref{1Deffhamilt}) and that the
commutation relations Eqs.~(\ref{anomaly}) are valid for the case of
$\nu<1$ also. An exception occurs for large $m$ when the ground state
of the 2D ES is expected\cite{wigcrys} to be a Wigner crystal.

The relation of the phase fields to the densities of left going and
right going fermions is given in reciprocal space 
\begin{mathletters}
\begin{eqnarray}
\theta_q &=& \sqrt{\frac{\pi}{\nu}} \frac{i}{q} \, \left[
\varrho^{\mbox{\tiny R}}_q + \varrho^{\mbox{\tiny L}}_q \right] \\
\phi_q &=& \sqrt{\frac{\pi}{\nu}} \frac{i}{q} \, \left[
\varrho^{\mbox{\tiny R}}_q - \varrho^{\mbox{\tiny L}}_q \right]
\end{eqnarray}
\end{mathletters}
We use reciprocal space language for the following discussion because
it is convenient for dealing with anomalies caused by the long-range
of the electron-electron interaction. The Hamiltonian
(\ref{1Deffhamilt}) is quadratic in the phase fields:
\begin{equation}\label{phaseham}
H = \frac{1}{2 L} \sum_{q} \, |q| \, E_q \left\{ \frac{1}{g_q}
| \theta_q |^2 + g_q \, | \phi_q |^2 \right\}. 
\end{equation}
Here $E_q$ is the energy of the low-lying bosonic excitations of the
system, generally referred to as plasmons in the quantum wire case
and as edge magnetoplasmons in a QH bar. The dispersion relation
$E_q$ for the plasmon excitations reads
\begin{equation}\label{disper}
E_q = |q| \, \left[ \hbar v_{\mbox{\tiny F}} + \frac{\nu }{2 \pi}
V^{\mbox{\tiny intra}}_q \right] \sqrt{ (1 + \xi_q) (1 - \xi_q) }
\end{equation}
and the interaction parameter $g_q$ is defined as
\begin{equation}\label{gfactor}
g_q = \sqrt{ \frac{1 - \xi_q}{1 + \xi_q}}\mbox{ .}
\end{equation}
The parameter $\xi_q$ measures the relative strengths of inter- and
intra-edge contributions to the plasmon energy, and its formal
expression is
\begin{equation}\label{interintra}
\xi_q := \frac{V^{\mbox{\tiny inter}}_q}{V^{\mbox{\tiny intra}}_q +
2 \pi \hbar v_{\mbox{\tiny F}} / \nu} \mbox{ .}
\end{equation}
Obviously, $\xi_q = 0$ in the absence of inter-edge interaction. For
larger $v_{\mbox{\tiny F}}$, {\em i.e.\ } sharper confining potential,
$\xi_q$ is smaller, and we expect corrections due to
$V^{\mbox{\tiny inter}}_q$ to be less important. Expressions
(\ref{disper}) and (\ref{interintra}) reflect the interchangeability
of one-body and two-body contributions to the plasmon dispersion and
$V^{\mbox{\tiny intra}}_q$. Note that in samples with aspect ratios
close to unity the value of $g_q$ is close to $1$ even at the
smallest physically-relevant values of $q \sim L^{-1}$. Using these
results we can now compare the physical properties of excitations in
quantum wire and QH bar systems.
\subsection{Plasmon Dispersion}
In order to do detailed calculations, we must specify the confining
potential of the QH bar. In what follows, we adopt the neutralizing
background model which produces sharp confinement and diminishes
corrections due to the inter-edge interaction as explained above.
To calculate the Fermi velocity resulting from a uniform neutralizing
background, we can use Eq.\ (\ref{elstatic}) apart from a sign change
since the Hartree energy of the Fermi sea is simply the electrostatic
potential due to the electron charge density in the ground state.
With the fractional filling factor incorporated properly, we find
that 
\begin{equation}
v_{\mbox{\tiny F}} = \nu \, \frac{e^2}{\hbar \epsilon \pi}\,
\ln{\left[ \sqrt{2\gamma} \frac{W}{\ell} \right]} \mbox{ .}
\end{equation}
For a macroscopic QH sample, $W / \ell$ is rather large (hence
$v_{\mbox{\tiny F}}$ is big), and $W \sim L$. Therefore, the bare
Fermi velocity in this system is larger than the renormalization
terms arising from interaction effects ($V^{\mbox{\tiny intra}}_q$
and $V^{\mbox{\tiny inter}}_q$). It is interesting to compare the 
$W \sim L$ limit with the case of a perfectly circular quantum
dot.\cite{ahm:aust:93,pit:prl:94,bla:int:95} In that case it follows
purely from symmetry arguments that in the long-wavelength limit the
plasmon energy is determined completely by the external potential
term in the Hamiltonian. We expect this to be approximately true for
all samples with aspect ratios close to one. The plasmon dispersion
in this limit is
\begin{table}
\caption{Contributions to the plasmon dispersion in quantum wire and
quantum-Hall (QH) bar systems.  This Table is based on the common
separation of energies in electronic systems into kinetic, Hartree,
external potential, and exchange-correlation contributions. Table
entries indicate whether the energy mentioned in the left column
gives a positive ($+$), negative ($-$), or zero ($0$) contribution to
the energy of long wavelength plasmons.}
\begin{tabular}{ccc}
 & quantum wire & QH bar \\ \hline
kinetic energy & $+$ & $0$ \\
external confining potential & $0$ & $+$ \\
Hartree energy & $0$ & $-$ \\
exchange/correlation energy & $+$ & $+$ \\ \hline
\end{tabular}
\label{compare}
\end{table}
\begin{equation}\label{macrodisperse}
E_q = - \nu \, \frac{e^2}{\epsilon \pi} \, |q| \, \ln{\left[ 
\sqrt{\gamma /8} \,\, |q| \ell \right]}\mbox{ .}
\end{equation}
Expressions of the form of Eq.~(\ref{macrodisperse}) for the edge
magnetoplasmon dispersion have been successfully applied to interpret
experimental data\cite{hls:prb:83,tal:jetp:86,heit:prl:90,wass:prb:90,
heit:prl:91,ray:prb:92,zhi:prl:93,hel1:85,hel2:85,hel:91} and were
originally derived classically\cite{vav-sam:jetp:88,bla:phyb:92}.
That Eq.~(\ref{macrodisperse}) is the correct result for
experimentally-realistic QH samples is one of the main points of our
discussion here. This result should be contrasted with the plasmon
dispersion relation at long wavelengths in a quantum wire where (see
also Ref.~\onlinecite{brey:95}) it is found\cite{dassarma} that $E_q
\sim |q| \sqrt{|\ln [|q|]|}$. 
More generally the underlying microscopic terms in the Hamiltonian
differ qualitatively in their influence on the magnetoplasmon
dispersion in the two cases as summarized in Table \ref{compare}
which is based on the common separation of energies into kinetic,
Hartree, external potential, and exchange-correlation contributions.  
\subsection{Correlation Functions}
The phase field formalism facilitates the straightforward calculation
of electronic correlation functions. A hallmark of the bosonization
approach in the study of 1D
systems\cite{emery,sol:adv:79,fdmh:jpc:81} is the possibility to
express the electron field operator $\psi$ in terms of the phase
fields. Electron Green's functions can then be written in terms of
Green's functions of the phase fields, which are readily calculated
because the Hamiltonian (\ref{phaseham}) is quadratic.

In truly 1D systems, the electron field operator depends on {\em one}
spatial coordinate ($x$) only. Incorporating the 2D aspect of a
quasi-1D ES and, in particular, a QH bar, we have to consider its
dependence on the lateral coordinate ($y$) as well:
\begin{equation}\label{electron}
\psi_{\mbox{\tiny L,R}}(x, y) = \Phi_{\mbox{\tiny L,R}}(x, y) \exp{
\left[ \pm i \sqrt{\frac{\pi}{\nu}} \theta(x) + i
\sqrt{\frac{\pi}{\nu}} \phi(x) \right] }
\end{equation}
where $\Phi^{\mbox{\tiny QW}}_{\mbox{\tiny L,R}}(x, y) :=
\psi^{\mbox{\tiny QW}}_{0, \mp k_F} (x, y)$ for the quantum wire and 
$\Phi^{\mbox{\tiny HB}}_{\mbox{\tiny L,R}}(x, y) :=
\psi^{\mbox{\tiny HB}}_{0, \mp k_F} (x, y)$ for the QH bar. The
operator for the total density of electrons at some position along a
quantum Hall bar can be obtained by integrating over the transverse
($y$) coordinate:
\begin{equation}\label{totaldensity}
\varrho(x) = \int dy \left[ \psi_{\mbox{\tiny L}}(x,y)
+\psi_{\mbox{\tiny R}}(x,y)\right]^{\dagger} \left[
\psi_{\mbox{\tiny L}}(x,y) +\psi_{\mbox{\tiny R}}(x,y)\right]\mbox{ .}
\end{equation}
In terms of the phase fields we find that 
\begin{mathletters}
\begin{equation}\label{osccharge}
\varrho(x) = \sqrt{\frac{\nu}{\pi}} \partial_x \theta(x) + 
\hat O_{\mbox{\tiny CDW}} (x) 
\end{equation}
where $\hat O_{\mbox{\tiny CDW}}$ represents the portion of
the charge density which oscillates with period $\pi / k_F $.
The slow decay of correlations associated with this part of the
charge density\cite{hjs:prl:93} is the basis of the quasi-crystalline
character of the electrons in a quantum wire. The spatial separation
of oppositely directed electrons in quantum Hall bars is not
important in the first term of Eq.~(\ref{osccharge}) because the
$y$-dependent part of $\Phi_{\mbox{\tiny L,R}}(x, y)$ is normalized.
However in the $2 k_F$ term, the overlap of $\chi_0(y)$'s at
$\pm k_F$ enters and the spatial separation changes the result. We
find that 
\begin{equation}
\hat O_{\mbox{\tiny CDW}}(x) =  \frac{1}{\pi \ell} \exp{\left[ \left(
-\frac{W}{2\ell} \right)^2 \right]} \cos{\left[2 k_F x + 2
\sqrt{\frac{\pi}{\nu}} \theta(x) \right]} \mbox{ .}
\end{equation}
\end{mathletters}
The quantum wire case can be recovered by setting $W =0$. In the Hall
bar case $\hat O_{\mbox{\tiny CDW}}$ acquires the prefactor $\exp{[
- (W/\ell)^2 ]}$ which, for realistic sample dimensions ($W/\ell \sim
10\dots 100$), is extremely small. 

\begin{figure}[b]
\epsfxsize3.6in
\centerline{\epsffile{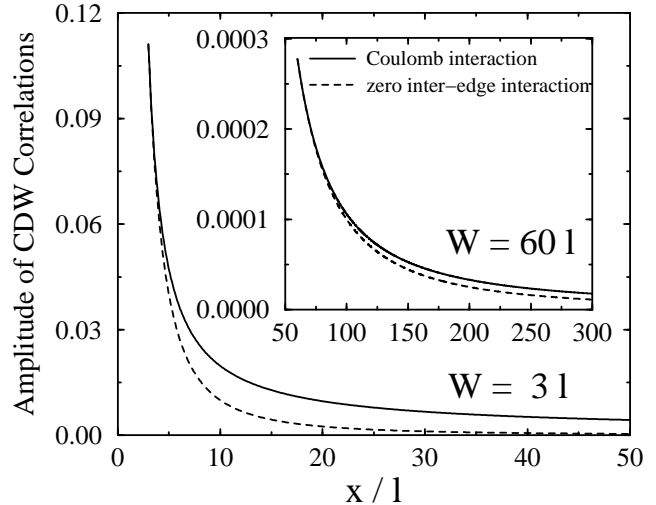}}
\caption{CDW-correlations at the edge of a QH bar. Plotted as the
solid line is the $x$-dependent exponential on the r.h.s.\ of
Eq.~(\ref{cdwcorr}) at fixed $W / \ell$ for filling factor $\nu = 1$.
The result $\left(x/\ell \right)^{-2/\nu}$ (which one would obtain if
inter-edge interaction was not present) is shown as the dotted line.
For large $W$, the difference between the two curves is small for $x$
less than several times $W$. In samples with large aspect ratios, the
correlation functions have a large relative difference for $x \gg W$
but both are already quite small in magnitude. The quasi-solid
correlations in quantum wires are therefore unlikely to be of
physical importance in typical Hall bar samples.}
\label{cdwfig}
\end{figure}

Using the bosonization technique, we find for the CDW correlations
\begin{eqnarray}
&&\left< \hat O_{\mbox{\tiny CDW}}(x) \, \hat O_{\mbox{\tiny CDW}}(0)
\right> = \frac{1}{2\pi^2 \ell^2} \exp{\left[-\left(W/\ell\right)^2
\right]} \cos{(2 k_F x)} \nonumber \\ && \hspace{2.5cm} \times \exp{
\left\{ -\frac{2\pi}{\nu} \left< \left[ \theta(x) - \theta(0)
\right]^2 \right> \right\} } \mbox{ .}
\end{eqnarray}
The phase field correlation function is determined by the interaction
parameter $g_q$ which reflects the non-Fermi-liquid properties of the
system\cite{emery,sol:adv:79,fdmh:jpc:81} resulting from inter-edge
interaction. A standard calculation\cite{hjs:prl:93} gives the
following result
\begin{equation}
\left< \left[ \theta(x) - \theta(0) \right]^2 \right> = 
\frac{2}{L}\sum_{q > 0}\,\, g_q\,\frac{1 - \cos{(q x)}}{q} \mbox{ .}
\end{equation}
For the case of the Coulomb interaction we find\cite{hjs:prl:93}
(for $x\geq W$) that
\begin{eqnarray}\label{cdwcorr}
&& \left< \hat O_{\mbox{\tiny CDW}}(x) \, \hat O_{\mbox{\tiny CDW}}
(0) \right> = \frac{1}{2\pi^2 \ell^2} \exp{\left[-\left(W/\ell
\right)^2\right]} \cos{(2 k_F x)} \nonumber \\ && \hspace{2cm} \times
\exp{\left[ -\frac{2}{\nu} \sqrt{ \left| \ln[W/\ell] \, \ln [x^2/(W
\ell)] \right|} \right]} \mbox{ .}
\end{eqnarray}
The very weak dependence on $x$ at $x \gg W$ is associated with the
vanishing of $g_q$ for $qW \ll 1$. The factor $\exp{\left[ -
\frac{2}{\nu} \sqrt{ \left| \ln[W/\ell] \, \ln [x^2/(W \ell)] 
\right|} \right]}$ on the r.h.s.\ of Eq.~(\ref{cdwcorr}) is plotted
as a function of $x$ for two different values of $W$ in
Fig.~\ref{cdwfig}. For comparison, the power-law expected when
inter-edge interactions are absent
\begin{equation}\label{powerlaw}
\left< \hat O_{\mbox{\tiny CDW}}(x) \, \hat O_{\mbox{\tiny CDW}}(0)
\right> \sim \left(\frac{x}{\ell}\right)^{-2/ \nu} \mbox{ if }
V^{\mbox{\tiny inter}}_q \equiv 0
\end{equation}
is also plotted. The correlations shown in Eq.~(\ref{cdwcorr}) fall
off more slowly than they would if inter-edge interactions were
neglected. However, for large $W/\ell$, which is the case applicable
to QH samples, the relative difference between Eq.~(\ref{cdwcorr})
and the power-law limit [Eq.~(\ref{powerlaw})] remains small even
when $x$ is several times larger than $W$, see Fig.~\ref{cdwfig}.
For Hall bar samples with aspect ratios $L/W$ of order less than
$\sim 10$, even the minimum value of $g_q$ (for $q=2\pi/L$) is close
to unity. Therefore, there is no regime where quasi-solid
correlations occur.

\section{Summary}
In conclusion, we have calculated microscopically the dispersion
relation for edge magnetoplasmons in a QH bar, emphasizing 
distinctions between these excitations and the plasmon excitations
of a quantum wire. We have carefully examined the influence of
different terms in the total Hamiltonian on plasmon excitations and
obtain quite different results in Hall bar and quantum wire cases.
Whereas the plasmon energy in the quantum wire case has contributions
only from kinetic energy gain and exchange energy loss in the
underlying electron system, the energy of edge magnetoplasmons in a
QH bar has additional contributions arising from electrostatic
(Hartree) and external potential terms but no contribution due to
kinetic energy gain. Despite these differences, the low-energy
effective Hamiltonian for both systems is identical for $q W \ll 1$.

Typical sample geometries of QH bars have an aspect ratio close to
unity. Therefore, in this case, $q W \geq 1$ even for the smallest
possible $q$. In this limit, inter-edge interaction is negligibly
small. As a result we find that for typical sample geometries, the
classical magnetoplasmon dispersion relation,
Eq.~(\ref{macrodisperse}), which differs from the plasmon dispersion
relation for a quantum wire, is accurate. Important corrections to
the plasmon dispersion due to the interaction between left and right
moving parts of the electron density occur only for long narrow Hall
bar samples. We also find that typical QH samples are not in the
regime where the quasi-solid behavior expected for charge density
correlations in a quantum wire is important. Furthermore, the spatial
separation of left movers and right movers in a QH bar leads to a
suppression of the CDW-fluctuations which is a Gaussian function
of the width $W$ of the Hall bar. 

\acknowledgements
The authors acknowledge many helpful discussions with S.~M.\ Girvin,
J.~J.\ Palacios, and R.\ Haussmann. This work was funded in part by
NSF grant DMR-9416906. U.Z.\ thanks Studienstiftung des deutschen
Volkes (Bonn, Germany) for financial support.


\begin{thebibliography}{10}
\bibitem{caveatsf} In this paper we assume that the field is
strong enough that $m^* \omega_c^2$ always exceeds the 
curvature of the confining potential.
\end{thebibliography}
\end{document}